# Confidence intervals for normalised citation counts: Can they delimit underlying research capability?[1]

Mike Thelwall, Statistical Cybermetrics Research Group, University of Wolverhampton, UK

Normalised citation counts are routinely used to assess the average impact of research groups or nations. There is controversy over whether confidence intervals for them are theoretically valid or practically useful. In response, this article introduces the concept of a group's underlying research capability to produce impactful research. It then investigates whether confidence intervals could delimit the underlying capability of a group in practice. From 123120 confidence interval comparisons for the average citation impact of the national outputs of ten countries within 36 individual large monodisciplinary journals, moderately fewer than 95% of subsequent indicator values fall within 95% confidence intervals from prior years, with the percentage declining over time. This is consistent with confidence intervals effectively delimiting the research capability of a group, although it does not prove that this is the cause of the results. The results are unaffected by whether internationally collaborative articles are included.

## 1. Introduction

The average citation impact of the publications of a group of researchers can be calculated with various field normalised indicators (Waltman, van Eck, van Leeuwen, Visser, & van Raan, 2011ab). The results may inform funding decisions for departments or higher level policy-making (e.g., Science-Metrix, 2015). The underlying assumption is that the number of citations to a scholarly paper tends to reflect its contribution to science (Merton, 1973). Although this is frequently untrue (MacRoberts & MacRoberts, 1996) or an oversimplification for individual citations and citation counts (Borgman & Furner, 2002), it is reasonable when applied to sufficiently large groups of papers in many disciplines (van Raan, 1998).

Citation counts need to be normalised for the field and year of publication to avoid misleading results. Nevertheless, even normalised citation indicators do not accurately compute the impact of a group of publications because field and year normalisation is unavoidably imperfect. These imperfections include treating articles that are published on different days during the year as being published at the same time (giving earlier articles an advantage) and combining interdisciplinary research and related specialisms into "fields" for normalisation purposes or using another heuristic to define fields (giving articles from higher citation specialisms or multidisciplinary combinations an advantage). Moreover, citations do not reflect all types of contributions to scholarship or non-scholarly impacts.

This article investigates whether it is reasonable to use confidence intervals to delimit the likely underlying research capability of a group. It does this by first calculating 95% confidence intervals for indicators annually. It then checks how often indicator values for a group fall within 95% confidence intervals from previous years. Finally, it heuristically assesses whether this frequency is broadly in line with the hypothesis that the confidence intervals delimit underlying research capability. This assessment necessarily relies upon subjective assumptions about the level of stability to be expected in different contexts.

## 2. Confidence intervals and random factors for citation-based indicators

A controversy exists over whether confidence intervals should be reported alongside field normalised indicators (Williams & Bornmann, 2016). For example, stability intervals have been suggested as a partial substitute, using bootstrapping to assess the effect of changes in underlying publication dataset on indicator values (Waltman, Calero-Medina, Kosten, Noyons, Tijssen, et al., 2012). Citation-based evaluations usually incorporate all publications produced by the assessed group (e.g., department, institution, nation) within the given period or all those that meet a criterion (e.g., indexed in Scopus). They are therefore not a random sample of a larger population but a complete set or specified subset (Schneider, 2016). Thus, from a basic statistical perspective it does not make sense to calculate confidence intervals for them because confidence intervals are designed to help infer population parameters from a random sample. In other words, citation indicators calculated from complete publication sets are exact rather than estimates.

Despite the above logic, there are reasons why confidence intervals can be desirable. There are two main types of random factors that may influence the number of citations accrued by the publications of a research group: citation-related and article-related. Many citation-related factors are out of the control of the authors of a paper (Abramo, Costa, & D'Angelo, 2015; Waltman, 2016). Each individual citation to a paper in the Web of Science (WoS) or Scopus is a consequence of multiple factors, including someone's decision to write a new paper, the new authors finding the first paper and judging it worth citing, the new authors submitting their paper to a WoS/Scopus journal, and the journal's editors and referees agreeing to accept the paper. Thus, if a research group's outputs receive a total of 1000 citations then this outcome is the result of thousands of individual decisions. From this perspective, confidence intervals for normalised citation counts would be reasonable to assess the range of values that the research group *could* plausibly have achieved after publishing their papers. Here, the confidence interval is for factors external to the research group that affect their citations. The set of citation counts is treated as an apparent population (Berk, Western, & Weiss, 1995): a sample of the outcomes that might reasonably have been expected from the set of publications. Although randomness at the citation level is a relevant phenomenon, it is not the main focus of the current paper. Instead, the type of randomness primarily modelled in the current paper is at the publication level.

At the publication level, each output from a group is a consequence of the creative powers of its authors as well as their technical prowess and the availability of time and other resources to conduct the research. Thus, each group publication is partly due to creativity-related factors that are not fully within the authors' control (Lee, Walsh, & Wang, 2015; Simonton, 2004). Whilst a highly creative author may *tend* to produce high impact work, she is not able to *guarantee* that all her ideas have the same high impact. Instead, for reasons that she may not fully understand or control, some of her papers may remain uncited whereas others become citation classics. For example, the Google Scholar citation profile of Nobel Prize winner Michael Levitt includes articles with citation counts ranging from 0 to 3,572. This level of variability seems too great to be explained by a citation-level random factors model.

In theory, creativity-related factors could be conceived and modelled as each person having an underlying research-related creative power but each of their publications randomly varying above or below this value. This explains why the same group of authors can produce works with substantially different research impacts (however assessed).

## 2.1 The research capability model

If the above publication-level random factors logic is accepted then a confidence interval for an impact indicator partly reflects the underlying ability of a group to produce impactful work in addition to external, citation-related factors. From this perspective, the set of publications produced by a group is not a definitive finite population (Nane, 2016) but is a sample from an infinite set of the publications that the group might have written in similar circumstances (Claveau, 2016; Williams & Bornmann, 2016). This publication-level randomness is primarily modelled in the current paper and the theoretical background introduced here will be called the *research capability model*. For clarity, it can be distinguished from two other perspectives.

- *The research capability model*: A group of researchers (of any size, including a single person) has a fixed underlying capability to produce research of a given quality or impact (or range of qualities) but the quality and impact of that research is affected by factors outside their control, including creativity variability, field changes, and mission changes. This "fixed underlying capability" may have distributional characteristics, such as 50% of the group being able to produce excellent research and 50% being able to produce average research. From a research capability perspective, citation analyses would aim to estimate the underlying research capability of a research group from the citations to its publications. This analysis would need to consider both publication-level and citation-level random factors.
    - For example, if group A had a field normalised citation score of 1.1 and group B had a field normalised citation score of 1.2 then statistical inference would be needed to assess the likelihood that group B had a higher underlying research capability than group A.
- *The research achievement model*: Irrespective of whether they have an underlying research capability, a group of researchers produces a set of *publications* that has an underlying quality or impact. From a research achievement perspective, citation analysis would aim to estimate the underlying quality or impact of the publications produced from the citations to them. This analysis would only need to consider citation-level random factors. Here the focus is on the quality of the research that the group produced (i.e., their achievements), irrespective of what they were capable of producing.
    - If group A had a field normalised citation score of 1.1 and group B had a field normalised citation score of 1.2 then statistical inference would be needed to assess the likelihood that group B had produced higher impact or quality publications than group A.
- *The impact recording model*: Irrespective of whether they have an underlying research capability or the average quality of their publications, a group of researchers produces a set of publications that have citation counts in a citation index and these are equated with their value. From an impact recording perspective, citation analysis would aim to report accurate citation-based indicators for their publications. This analysis would not consider any random factors but would interpret all data at face value.

- If group A had a field normalised citation score of 1.1 and group B had a field normalised citation score of 1.2 then it would be concluded that group B had, on average, generated higher average impact than group A.

If the research capability model assumptions are accepted, then inferential statistics or confidence intervals can help decide whether one group has a greater capability to produce high impact work than another (Claveau, 2016; Williams & Bornmann, 2016). The equivalent is true for the research achievement model. If the difference between two groups is small, as judged by confidence intervals, then it is reasonably likely that they have similar underlying capabilities to produce impactful research and that the indicator differences found reflect minor uncontrollable internal and external variations in the outputs produced in each period. Similarly, confidence intervals can be used to judge whether differences over time are statistically significant. Governments and universities regularly introduce research policy changes and assess the influence of these policy changes on research impact. For this, confidence intervals would help to reveal whether the new policy had significantly changed the underlying ability of the groups or whether the inevitable differences were too small to indicate genuine underlying change.

The formerly proposed stability intervals match the research capability model because they are based on random changes in the underlying publication dataset (Waltman, Calero-Medina, Kosten, Noyons, Tijssen, et al., 2012).

## 2.2 Theoretical limitations of confidence intervals

Confidence intervals may be called upon to address the statistical inference needs of the research capability or research achievement models above. All statistical techniques currently used in citation analysis applications do not seem to distinguish between the two sets of assumptions (i.e., publication-level and citation-level random factors or just citation-level random factors) and so the difference would be in the interpretation of the results rather than in the calculations. It is difficult to distinguish between publication-level and citation-level random factors because this would require theoretical assumptions about the likely distribution of publication impacts (or citations) to split citation counts into publication-level and citation-level components (and perhaps other factors: e.g., Dieks & Chang, 1976).

An important limitation with confidence intervals is that they are based upon the assumption that the citation count for each article is independent of the citation counts of all other articles, and only dependant on random factors and the capability of the producing group, which is untrue (Schneider, 2013). For example, a group may publish a set of high impact articles on a new topic for which they enjoy an early research advantage, or may be constrained to produce many low impact publications for an externally funded project. In both cases a single cause underlies an increase or decrease in citation impact for multiple articles. Since on a large scale these factors cannot be quantified, confidence intervals lack statistical validity and can only be estimates. Whilst there are methods to calculate confidence intervals for variables that are not independent (Dunn, 1959; Goldstein, 2011), these require information about the nature of the dependence, such as the groups of articles that are related to each other, which cannot fully be known.

Since, as argued above, confidence intervals can be helpful in research evaluation but due to independence issues lack a theoretical basis to validate their use in practice, experiments with real data are needed to assess whether they tend to give actionable information.

## 3. Underlying research capacity stability

This article addresses the assumptions described above about confidence intervals for citation-based indicators under the research capability model. To empirically assess the limitations of confidence intervals to delimit underlying research capability, data would be needed for the same group of researchers producing publications under identical conditions during two time periods. For field normalised indictors it is not only necessary that the group assessed does not change but also that the rest of the world does not change because the citation counts for the rest of the world are needed for the normalisation calculation. Since the conditions for research change continually, the only way in which this could be achieved would be to take two samples from the same period by splitting all publications from that period into two at random. Given the known distribution of citation counts (Eom & Fortunato, 2011; Evans, Kaube, & Hopkins, 2012; Thelwall & Wilson, 2014; Thelwall, 2016ab) and the assumptions used to generate the confidence interval formulae in the first place, this would almost be the same as testing confidence intervals from theoretical populations (Thelwall & Fairclough, 2017) and so would be uninformative. More seriously, *it would not test the key unknown quantity of confidence intervals: the degree to which citation counts are not independent of each other*. In contrast, the output of a group (e.g., country) a year later may be partly due to non-independent influences, such as personnel or topic changes.

### 3.1 The stability assumption

To address the above issue, this article makes an additional *stability assumption*: that the underlying research capability of a group (e.g., country) is unlikely to change much in the short term (e.g., 1 year) and more likely to change in the long term (e.g., 10 years). This assumption only makes sense under the research capability model and not the other two models in Section 2.1, which are only concerned with the impact of the publications produced during a given period.

In practice, research indicators for countries or other large groups of researchers tend to be stable over time (e.g., Colliander & Ahlgren, 2011; for countries: Aksnes, Schneider, & Gunnarsson, 2012; Archambault, Campbell, Gingras, & Larivière, 2009; Elsevier, 2013; Fairclough & Thelwall, 2015), but can experience longer term substantial shifts in response to policy changes (Aagaard & Schneider, 2016; Liu, Tang, Gu, & Hu, 2015). It is therefore reasonable to compare indicators from similar periods of time, such as consecutive years, on the basis that the context will rarely change much.

### 3.2 The stability assumption and confidence intervals

If the stability assumption is valid, then the same group operating within the same conditions at two close points in time would attain indicator values that were within 95% confidence intervals moderately under 95% of the time, if the sample sizes were not too small (see Section 5.2). If no publication-level or citation-level random factors were present in the production and citation processes then under the same conditions, a group would produce an identical indicator score and so 95% confidence intervals calculated from the first period would contain the indicator value for the second period 100% of the time.

Conversely, if the impact of a group's outputs from one year were unrelated in any way to their outputs from the next year then confidence intervals would be useless and much fewer than 95% of subsequent year values would fall within the original 95% confidence limits, with no change over time.

## 4. Research questions

Following the above discussions, the first research question addresses the usefulness of confidence intervals to delimit the underlying research capability of a group (e.g., country) by applying the stability assumption and using empirical data to assess whether subsequent indicator values fall within confidence intervals from previous years plausibly often.

- When 95% confidence intervals are calculated for field-normalised indicators for a group of publications that is not very small, do moderately less than 95% of the same indicators calculated a short period later tend to be within the 95% confidence intervals?

With the stability assumption (Section 3.1), the term "moderately" will be interpreted in terms of the extent to which the proportion of times the subsequent indicators fall outside a 95% confidence interval in comparison to what would be expected in a situation with no change, as estimated by the 0 years offset value in the graphs (calculated using the Section 5.3 bootstrapping method). The phrase "that is not very small" in the research question relates to the discussion in Section 5.2 and is left imprecise in the current paper.

Following the stability assumption, the longer the period between the original confidence interval calculations and the subsequent indicator value calculations, the greater the likely cumulative changes to the group and the world system and hence the lower the probability that the subsequent indicator value would fall within a 95% confidence interval. This leads to the second research question.

- When 95% confidence intervals are calculated for field-normalised indicators for a group of publications that is not very small, does the probability that the same indicators calculated later fall within the original 95% confidence intervals decrease steadily as the time interval increases?

## 5. Methods

To address the above research questions, a dataset is needed that is stable over time. A logical approach would be to take one or more Scopus or WoS subject categories and to calculate field normalised indictors for these at different points in time. This would introduce additional influences, however. Scopus and WoS subject categories can have anomalous inclusions (e.g., Abrizah, Zainab, Kiran, & Raj, 2013; for example, *Qualitative Health Research* is in the WoS set of Information Science & Library Science journals [and two other categories]) and are periodically updated in terms of the journals that fall within them. Thus, a single subject category can incorporate periodic shifts in focus or content due to changes in the constituent journals. There can also be technical issues with individual journals ceasing production, changing name or not being indexed for short periods that can lead to fluctuations in indicators. For example, a temporary absence of a high impact journal would deflate normalised indicator scores for groups that tended to publish in it, and inflate values for the remainder. Whilst these fluctuations are likely to be minor, they would compromise the validity of the tests reported here.

A purer strategy is to focus on individual journals. Even if some issues of a journal are lost, this would be unlikely to add a systematic bias because the lost data is likely to be comparable to the remaining data, unless it is a special issue. Focusing on a single journal also avoids subject classification problems. Nevertheless, journals change focus over time, either as a deliberate policy shift or due to editorial or publisher changes, but this seems likely to be a rare event. Large journals are needed to achieve reasonable statistical power and so large monodisciplinary journals indexed in Scopus were chosen. General journals were rejected (e.g., Science, PNAS) because effective field normalisation requires knowledge of the subject areas of all published articles.

Taking the above into consideration, the data source selected was all Scopus-indexed articles in the 36 monodisciplinary journals, as judged from their titles, within the 50 journals with the most articles indexed in Scopus in the start year, 1996 (Appendix). This data was downloaded from Scopus in February/March 2017. The citation counts for each article were normalised only against other articles from the same journal and year (following: Colliander & Ahlgren, 2011). Thus, the journal is treated as the field for the analysis, removing subject categorisation sources of fluctuations.

Stable groups of researchers were needed to assess changes over time. Individual research groups, departments or universities are probably not stable enough since departments and research groups may appear and close over time, making a substantial change in research capability within an institution. This can occur at the national level but will have a smaller effect unless it is part of a national programme of expansion or contraction within a specific field, which seems likely to be a much rarer event. To maximise statistical power, the largest countries were assessed, choosing the 10 countries with the most articles in the journals examined.

There are two main ways of assigning nationality to a journal article: whole and fractional counting (Van Hooydonk, 1997), although there are more complex methods (Hagen, 2010). Whole counting allocates a paper to a country if any of its authors have an affiliation to an institution within the country. Fractional counting allocates a k/n share of an article to a country if k out of n of its authors (or author affiliations) are from that country. Since the confidence interval formulae used here are not designed for fractional counting, two different whole counting approaches were used: inclusive and exclusive. Under inclusive counting, all articles with any author from the country were included in the set of papers from that country. Under exclusive counting, all articles with only authors from that country were counted (i.e.., excluding internationally co-authored articles). The latter case is better in theory because the results cannot be influenced by contributions from other countries. The former case is included for completeness because research collaboration is increasing (e.g., Adams, 2013; Brunson, Wang, & Laubenbacher, 2017; Puuska, Muhonen, & Leino, 2014) and is important to science (Gibbons, Limoges, Nowotny, Schwartzman, Scott, & Trow, 1994) and citation impact (Gazni, Larivière, & Didegah, 2016).

The period analysed was 1996 to 2014. The start year of 1996 was chosen since the coverage of Scopus expanded before this date. The end year 2014 allows over two years for article citation counts to mature. Although citation count maturation is not necessary for field normalised indicators, the lower average citation counts tend to reduce the magnitude of differences between countries and make indicators less stable.

## 5.1 Confidence interval formulae

There are many different field normalised indicators but the one used was the Mean Normalised Log-transformed Citation Score (MNLCS) (Thelwall, 2017a). It is a log-transformed variant of the Mean Normalised Citation Score (MNCS) (Waltman, van Eck, van Leeuwen, Visser, & van Raan, 2011ab). It was chosen for its confidence interval formulae (Thelwall, 2017b) and evidence about their accuracy (Thelwall & Fairclough, 2017). In this context, MNLCS is the ratio of the country to journal-wide mean of the log transformed citation counts $\ln(1 + c)$. The MNLCS value for a nation state $S$ (i.e., country) output in a journal $J$ and a single year is therefore:

$$MNLCS_{SJ} = \left. \frac{1}{|S|}\sum_{a \in S} \ln(1 + c_a) \middle/ \frac{1}{|J|}\sum_{a \in J} \ln(1 + c_a) \right. \quad (1)$$

Formula (4) below (from: Thelwall, 2017a) was used for 95% confidence intervals for the country MNLCS. This accounts for citation impact variability in the country and variability in the remainder of the journal, assuming neither are accurate (Feiller, 1954). In the formula, $SE_s$ and $SE_j$ are the standard errors and $\overline{c_s}$ and $\overline{c_j}$ are the arithmetic means of the normalised citations $\ln(1 + c)$ for the country and whole journal, respectively.

$$h = t_{n_1+n_2-2,\propto} \left(\frac{SE_J}{\overline{c_s}}\right)^2 \quad (2)$$

$$SE_{\text{MNLCS}_{SJ}} = \frac{\text{MNLCS}_{SJ}}{1-h}\sqrt{(1-h)\frac{SE_s^2}{\overline{c_s}^2} + \frac{SE_J^2}{\overline{c_J}^2}} \quad (3)$$

$$\left(\frac{\text{MNLCS}_{SJ}}{1-h} - t_{n_1+n_2-2,\propto}SE_{\text{MNLCS}_{SJ}}, \frac{\text{MNLCS}_{SJ}}{1-h} + t_{n_1+n_2-2,\propto}SE_{\text{MNLCS}_{SJ}}\right) \quad (4)$$

Under the assumptions of the current paper, these confidence intervals are for the underlying capability of a group (e.g., country) to produce impactful research. In other words, the confidence intervals are for the expected (MNLCS) citation impact of papers produced by a group under similar conditions.

## 5.2 Confidence interval formulae and indicator values for subsequent years

If confidence intervals are to be used to assess whether the underlying capacity of a group (e.g., country) has changed over time (rather than to estimate underlying research capability) then it is important to know the likelihood that indicator values from subsequent years fall within confidence intervals for the first year. This section addresses the issue of the frequency with which an indicator value calculated in year $n + 1$ can be expected to fall within a 95% confidence interval for the indicator calculated in year $n$ if the research capability and research environment do not change.

With this assumption, the citation counts for each article produced by the group can be modelled from the discretised lognormal distribution (Thelwall, 2016b) with the same two population parameters $ln\ddot{\mathcal{N}}(\mu, \sigma^2)$. Although the modelling is at the level of articles, the random factors could be due to article-level, citation-level random factors, or a combination of both. For the MNLCS indicator calculations, the citation counts are transformed with $ln(1 + x)$ and divided by the field (in this case journal) average for the year. The data used to calculate MNLCS values is therefore a set of normalised citation counts, each of which is drawn from the same ratio of log-transformed discretised lognormal distributions.

Since for any sample, the MNLCS denominator is fixed (the field average) and only the numerator varies (the log normalised citation counts) and the denominator (all articles in the field) has a large sample size relative to the numerator (all articles in the field published by a country), the denominator will be ignored to simplify the situation.

Consider also the further simplification that the MNLCS numerator values $ln(1 + x)$ are close to following the (continuous) normal distribution $\mathcal{N}(\mu_0, \sigma_0^2)$. Then the issue is the percentage of times which the sample mean of a random sample from $\mathcal{N}(\mu_0, \sigma_0^2)$ lies within a 95% confidence interval calculated from a second random sample from $\mathcal{N}(\mu_0, \sigma_0^2)$. If the first random sample is small and the second sample is large, then the second sample mean will be very close to $\mu_0$ and the percentage of times that the second sample mean is in the first sample 95% confidence interval for $\mu_0$ will be close to that for $\mu_0$, which is 95% by definition.

In contrast, suppose for simplicity that the first random sample is large and the second is the smallest possible, at 1. Then the sample statistics for the first random sample would be close to the population parameters and the 95% confidence interval for $\mu_0$ would be approximately $\mu_0 \pm 1.96\sigma_0/\sqrt{n}$. Since the second sample size is 1, the second sample mean is the same as the sole second sample member, taken from $\mathcal{N}(\mu_0, \sigma_0^2)$. As $n$ gets larger, the range $\mu_0 \pm 1.96\sigma_0/\sqrt{n}$ tends to $\mu_0 \pm 0$ and the probability that the second sample mean (=second sample member) is within $\mu_0 \pm 1.96\sigma_0/\sqrt{n}$ tends to 0, which is the probability that the one member of the second sample is exactly $\mu_0$.

In summary, depending on the relative sample sizes, the probability that a second sample mean falls within a 95% confidence interval for the population mean calculated from a first sample can take any value between 0 and 0.95. The exact value depends on the two sample sizes, and (unknown) population parameters $\mu_0$ and $\sigma_0$. Thus, it not straightforward to detect whether the underlying research capability of a group has changed from one period to another based on confidence intervals. Because of this and the unknown amount of change expected from year to year, the exact proportion below 0.95 of subsequent indicator values that fall within a prior confidence limit will not be calculated.

### *5.3 Bootstrap confidence intervals*

Following the above discussion, suppose that (as in the data for the current paper) the sample sizes are not large enough to assume that a second MNLCS value would have a probability close to 0.95 of being within a previous 95% MNLCS confidence interval for the underlying research capability. In this case, bootstrapping can be used to estimate the expected probability. This bootstrapping approach splits the data into two sets at random so that the situation for both samples is identical (i.e., the ideal situation described above) and then estimates, by replication, the percentage of times the MNLCS from a second random sample from the same year falls within a 95% confidence interval for the MNLCS calculated from the first sample. (This is the 0 years offset data in the graphs in Section 6.) This is therefore an estimate of the extent to which a second MNLCS would lie within a 95% confidence interval from a former year under conditions of no change.

Bootstrapping, as described here, is an imperfect estimation method because it halves the sample sizes. Nevertheless, there is no alternative, given that any formula based on values from the current year would give values close to 95% by construction and would therefore be uninformative.

## 5.4 Lag 0 estimates

In summary, the first stage of the research design was to calculate MNLCS field normalised citation impact indicators for ten countries in two different ways for the impact of their refereed journal articles in 36 large monodisciplinary journals published in each year from 1996 to 2014 (i.e., 36 x 10 x 19 = 6840), together with 95% confidence intervals. The second stage was to calculate the percentage of indicator values in each country, journal and year that were within the 95% confidence intervals for each previous year for which there was sufficient data to calculate a confidence interval.

As mentioned above, if all the theoretical assumptions underlying a statistical test are satisfied, then 95% confidence intervals for a parameter contain the population parameter 95% of the time. If two random samples are taken from the same distribution, then the sample parameter from the first set will fall within the 95% confidence interval from the second set less than 95% of the time because the first sample parameter is only an estimate of the population parameter. Nevertheless, other factors being equal, the larger the sample size for the first set, the closer to 95% is the chance of its sample parameter being within the second set 95% confidence interval. This approaches 95% for very large first set sample sizes. Thus, to identify temporal trends (see below), it is important to estimate the expected proportion of times the first sample MNLCS would fall within a 95% confidence interval for a second sample MNLCS.

For this (see Section 5.3), for each country, journal and year, confidence intervals were simulated for a lag of 0 by splitting each journal/year dataset into two (both for the journal and each individual country). MNLCS confidence intervals were calculated for one half of the data and then the second half was tested to see whether its MNLCS values for each country fitted within the confidence intervals. This contrived process halves the sample size and so is still an underestimate of the true percentage, with the underestimate being greater for smaller sample sizes.

## 5.5 Interpreting "moderately less" in the first research question

The first research question uses two related subjective terms, "moderately less" and "short period" and risks being true for all results by stretching their interpretation to fit the data and therefore being invalid. The purpose of the article is to assess whether it is plausible to conceive the publications as generated by an underlying stochastic process with a mean parameter (underlying research capability) that could be estimated by a formula and associated confidence interval. This plausibility is assessed subjectively by judging whether the proportion of means falling within the confidence interval over time is reasonable in the sense of being plausibly "moderate". The answer to the first research question would be negative if the proportion was above 95% or too far below 95% to be "plausibly" thought of as being due to "normal" (small) evolution over time.

To explain further, if the underlying research capability model was wrong then two extreme alternatives might be possible: (a) static - a research group's citations are an exact reflection of the group's capability and this research capability never changes so that all future indicator values are identical and fall within the confidence limits; or (b) random – groups do not have an underlying research capability but each year's average performance changes randomly from one year to the next (or, equivalently, research capability changes randomly between years) so the same low proportion of future indicator values fall within the confidence limits, irrespective of time. The task of this paper is to make a convincing subjective argument that the empirical trend observed, which ought to fall between these extremes, is plausible for the model being correct.

The purpose is *not* to give a confidence interval for next year's values because, under the assumption of the underlying research capability model, next year's values are also underlying research capability estimates. The confidence intervals give a 95% confidence for the underlying research capability of the group, rather than its future citation performance, although the two are connected through the model.

## 6. Results

For all countries and all year gaps examined, fewer than 95% of the indicator values for the country fall within 95% confidence intervals calculated from a previous year from one of the 36 journals, whether articles are considered with all (Figure 1) or any (Figure 2) authors from the country. This gives a positive answer to the first research question if the percentages for year 1 (72%-90% in Figure 1; 75%-92% in Figure 2) are accepted as being "moderately" below 95%.

Recall that the 0 years offset data in the graphs is different from the remaining data. It is a bootstrapped estimate of the percentage of times that a second (sample) MNLCS would fall within 95% confidence intervals for the underlying MNLCS in unchanging research conditions. Comparing the bootstrapped 0 years offset (recall that the sample sizes are half as large for the bootstrap sampling, so the Figure 1 values for 0 years are likely to be slight underestimates) with the real 1 year offset data, there is a sharp drop for several countries. Since country order for the year 1 offset in both graphs is approximately the size order in the dataset and the sharp drops do not occur for the largest countries, it is possible that this is a size effect. This is partly supported by previous evidence that research productivity varies more over time for smaller countries than for larger countries (e.g., for dentistry: Cartes-Velásquez & Delgado, 2014). The confidence intervals used to create the graph assume statistical independence between data values. More specifically, they assume that the number of citations received by each paper published by a country in a journal is unrelated to the number of citations received by any other paper published by that country in the journal, except for the citations that are due to the underlying research capability of the country. This is not true in practice, since researchers and research groups tend to produce sets of related papers and the arrival or departure of successful prolific researchers within the field of a journal can have an immediate effect on a country's indicator values unless it is large enough to absorb the difference. Thus, the tendency for values to fall outside of the 95% confidence interval more than expected could be due to several relatively sudden changes in research capability within the countries during 1996-2014. In summary, the first research question has a positive answer if the above factors can be accepted as plausibly accounting for the year 1 percentages (72%-90% in Figure 1; 75%-92% in Figure 2).

For all countries, the negative slope of the lines indicates a trend is for longer gaps to associate with a lower percentage of indicator values falling within a 95% confidence interval from the first year in the period examined. This gives a positive answer to the second research question (Figure 1, 2).

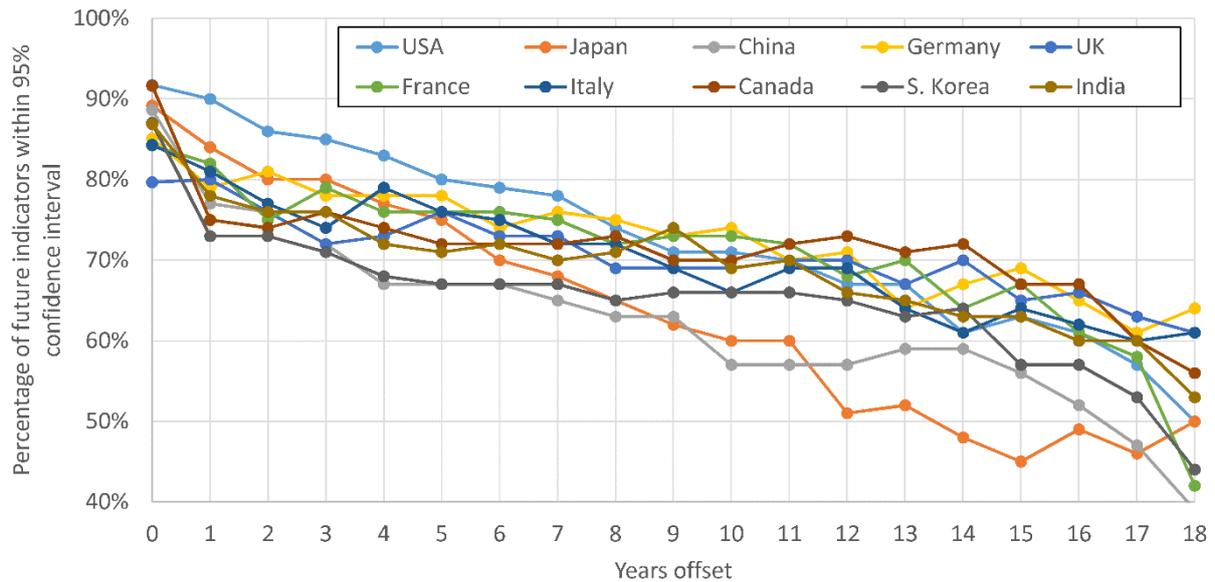

Figure 1. Percentage of MNLCS values within the 95% confidence intervals for the preceding year across the 36 journals by the number of years between the confidence interval and the subsequent indicator. Years run from 1996 to 2014. The data is for articles with all authors from the specified country. Sample sizes for individual points vary from 648 (1-year offset) to 36 (18 years offset) with an overall total of 123120 comparisons. The increased jaggedness on the right of the graph is due to smaller sample sizes. Countries are listed in decreasing order of number of articles in the whole dataset. Values for 0 years offset are from simulation on 50% of the data.

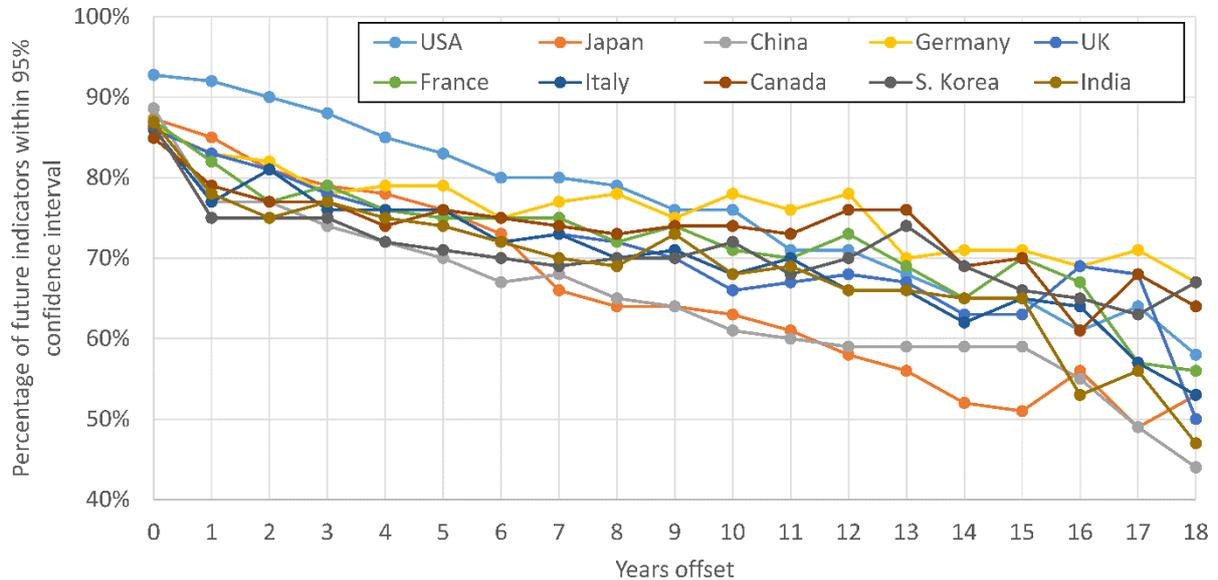

Figure 2. As for Figure 1 except that the data for articles with *any* author from the specified country (i.e. including internationally collaborative research).

To illustrate the process underlying the main graphs, Figure 3 plots one of the 360 country/journal combinations used in Figure 1, plus year 2015 and 2016 data. Here, the USA's indicator values are stable over time, but with a decreasing trend. The USA hovers around the world average (which is always 1 for the MNLCS) for this journal until about 2005 and then starts to systematically decrease. Once the decrease is large enough (2012), the indicator values tend to be outside of the 95% confidence intervals for most early years (up to 2005). This graph also reveals the anomalous behaviour after 2014, clarifying why the years 2015-2017 were not used in Figures 1 and 2.

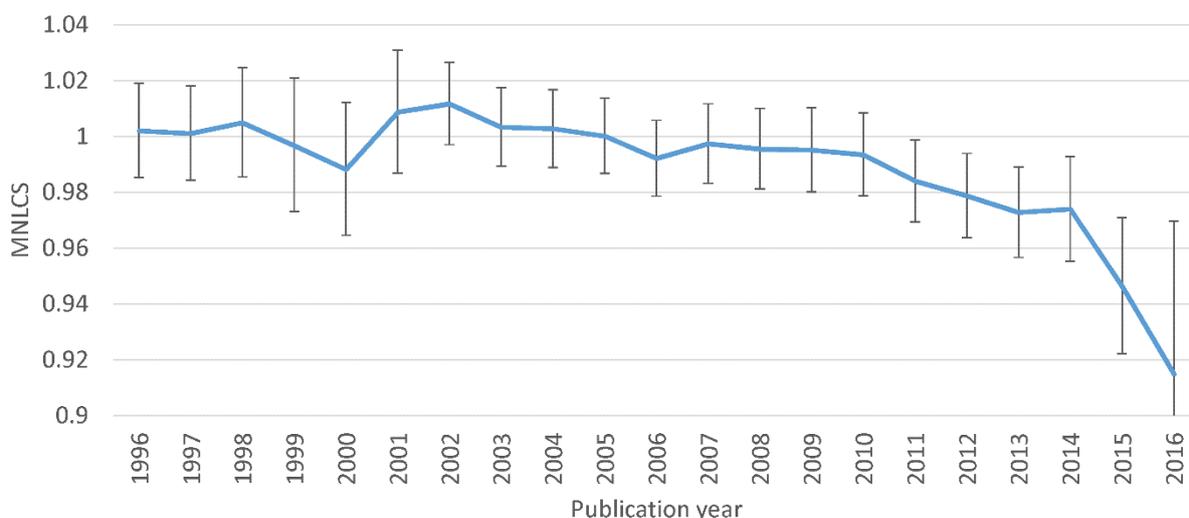

Figure 3. MNLCS and 95% confidence intervals for articles with all authors from the USA in the *Journal of the American Chemical Society*. Error bars show the 95% confidence intervals used in the calculation of Figure 1, with additional data for 2015 and 2016. This illustrates stable indicator values over time but with a slowly decreasing trend from about 2001.

Figure 4 is an example of a much less stable graph. The indicator values vary substantially from one year to the next, as can be seen from the jagged nature of Figure 4 compared to Figure 3, despite its larger y-axis span (0.8 compared to 0.14). The wider confidence intervals reflect the greater variability of the MNLCS average citation counts in this case. Taken together, both graphs show the value of the confidence interval approach for the task of predicting near future values and estimating the underlying research capability of a country. For the USA, policy makers should have much more confidence in the accuracy or stability of the indicator in Figure 3 than Japanese policy makers for Figure 4. For example, Japanese policy makers that ignored the confidence intervals and concluded in 1996 that their Astronomy & Astrophysics research capability was far below the world average might be greatly surprised to see that it had suddenly become better than the world average in 1997, unless a sudden major policy change had been implemented in 1996.

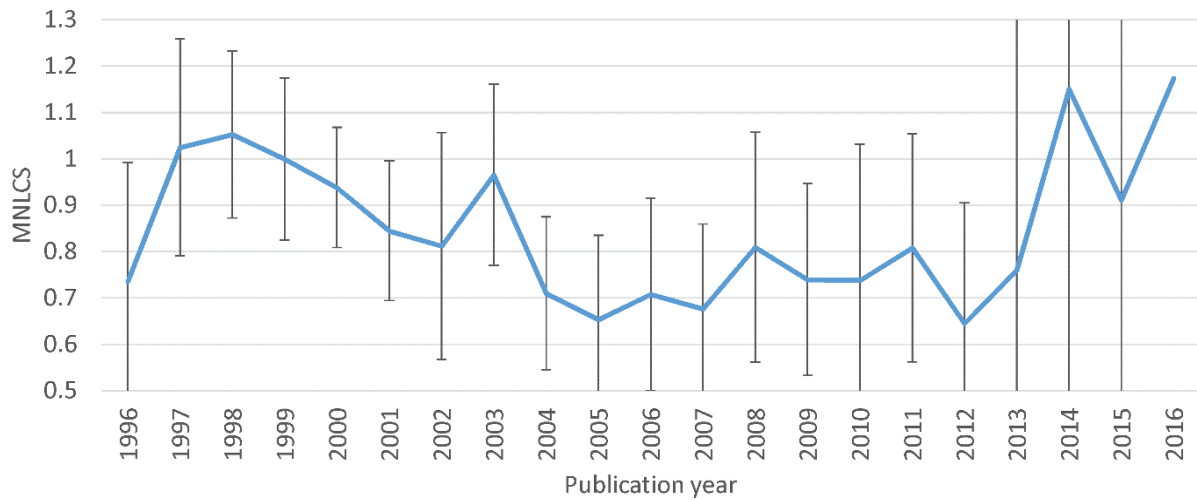

Figure 4. MNLCS and 95% confidence intervals for articles with any author from Japan in Astronomy & Astrophysics. Error bars show the 95% confidence intervals used in the calculation of Figure 2, with additional data for 2015 and 2016. This illustrates wide variations in values in parallel with wide confidence intervals. There is also a possible trend for lower research impact in the middle of the period.

## 7. Limitations

The conclusions drawn from this paper are limited by the artificial nature of the experiments conducted. The journal set covers mainly the life and natural sciences and excludes all social sciences and humanities. It would be rare for indicators to be constructed based upon individual journals rather than entire subject categories or sets of subject categories. This was necessary to give better conditions for comparing indicator values over time. In practice, indicators calculated from subject categories are likely to be more variable over time than those calculated here. In addition, the focus on large countries was also a practical one to give enough data to analyse.

For evaluations of individual departments or research groups based on the main subject categories that they publish in, the larger scope of the publications will partly compensate for the smaller numbers of researchers involved. This should not fundamentally affect the results since the confidence intervals incorporate sample size so that larger groups would tend to have smaller confidence intervals, other factors being equal.

The findings are also restricted to a single indicator for which confidence intervals have been proposed. Bootstrapping or formulae can be used to calculate confidence intervals for all other field normalised formulae but the accuracy of the results must be verified before they can be assessed over time, as in the experiment here.

The results suggest that the valid theoretical objection to confidence intervals based on the lack of independence of citation data (Schneider, 2013) tends not to be critical in practice: ignoring this objection will not cause inappropriate conclusions to be drawn by policy makers from confidence intervals. The reason why non-independence is not very important could be that it persists but that non-independent factors are stable over time rather than affecting a single year. For example, if a good researcher leaves a research group then they might recruit a similar one to replace it. Similarly, if an emerging topic researched by a group declines in importance, they might switch to another important topic over a few years, maintaining the average impact of their research.

Most fundamentally, the results do not prove the theoretical assumptions described above for confidence intervals: that they can delimit the normal range of variability in the outputs of a group of researchers, considering both internal sources of variation (e.g., creative variability) and external sources of variation (e.g., others' decisions affecting publication and citation). Instead, the results only give evidence that the use of confidence intervals in practice is broadly consistent with the assumptions underlying their use.

## 8. Conclusions

If the interpretation of "moderate" above is accepted then the results are broadly consistent with, but do not prove, the hypothesis that the citations accrued by the publications of a group (e.g., country) are partly due to random factors that are external (influencing others' decisions to cite the works) or internal (e.g., creativity-related factors accounting for natural variability in output impact) to the group. Field normalised MNLCS confidence intervals calculated from citation data contain the MNLCS value for the following year at about the same rate as would be expected if they were accurate estimates of the underlying capability of the group to produce citable research, and there were occasional sudden changes in the research group (personnel or topic changes) or environment (topic changes). It is therefore broadly reasonable to use appropriate confidence intervals to estimate the underlying research capability of a group in the sense of its capability to produce citable research. When used to compare different time periods, they are likely to be too narrow, however, due to internal and external personnel and topic changes. The use of confidence intervals should help to avoid decisions from being made based on insufficient evidence.

If policy makers are interested in the underlying research capability of a group, whether a department, university or country, then it is reasonable to delimit the likely ranges of this from the data using confidence intervals. MNLCS confidence intervals were used here but the same logic applies to other indicators with confidence limits, although the assumptions for these should also be tested to be safe.

In contrast, if administrators wish to reward a group for their performance over a given period (e.g., UK Research Excellence Framework: Murphy, 2017; http://www.ref.ac.uk/), then the decision is less clear cut. It would be reasonable to justify using the indicator value alone on the basis that this was what the group achieved in the assessment period and that the goal of the exercise is to reward achievement rather than to fund research capability. Although this would ignore the random external factors affecting citation counts, which is undesirable, it would also ignore the random internal factors, which is desirable in this context.

An important caveat for the use of confidence intervals is that their lay users may overinterpret them as delimiting the range of possible impacts from a group's research rather than its citation impact. Scientometricians should therefore be careful to emphasise that citation indicators only reflect one type of impact and are not designed to reflect the commercial, educational and societal impacts of a body of work, even if the two correlate (Terämä, Smallman, Lock, Johnson, & Austwick, 2016). In addition, statistical significance does not imply substantive importance and the independence assumptions underlying the confidence intervals are not valid (Schneider, 2013) and so confidence intervals should always be interpreted cautiously.

The data for the paper is available online https://doi.org/10.6084/m9.figshare.4910165.

## Appendix: Journals used

Journal of the American Chemical Society; Applied Physics Letters; Astronomy & Astrophysics; Astrophysical Journal; Biochemical & Biophysical Res. Comm.; Biochemistry; Brain Research; Cancer Research; Chemical Physics Letters; Inorganic Chemistry; Japanese J. of Applied Physics Part 1; J. of Agricultural & Food Chemistry; J. of Applied Physics; J. of Applied Polymer Science; J. of Biological Chemistry; J. of Chemical Physics; J. of Immunology; J. of Organic Chemistry; J. of Virology; Macromolecules; Physical Review Letters; Monthly Not. R. Astronomical Soc.; Tetrahedron; Tetrahedron Letters; Thin Solid Films; Geophysical Research Letters; Applied Mathematics & Computation; Nuclear Instruments & Meth. Physics A; Applied Surface Science; J. of Neuroscience; J. of Power Sources; Langmuir; Physica B Condensed Matter; Materials Science & Eng. A; Bioorganic & Medicinal Chemistry Lett.; Physical Review A.